\documentclass[aps,prb,twocolumn,showpacs]{revtex4}  
\usepackage{graphicx}  
\usepackage{dcolumn}   
\usepackage{bm}        
\usepackage{subscript}   
\usepackage{xcolor}  
\newcommand{\iac}[1]{{\color{black} #1}}
\newcommand{\jm}[1]{{\color{black} #1}}
\newcommand{\rc}[1]{{\color{black} #1}}

\begin{document}

\title{Immunity of intersubband polaritons to inhomogeneous broadening}
\author{J-M. Manceau$^{1,\dagger}$, G. Biasiol$^{2}$, N.L. Tran$^{1}$, I. Carusotto$^3$, R. Colombelli$^{1,\ddagger}$}
\address{$^1$Centre de Nanosciences et de Nanotechnologies, CNRS \jm{UMR 9001}, Univ. Paris-Sud, Universit\'e Paris-Saclay, C2N - Orsay, 91405 Orsay cedex, France\\
$^2$Laboratorio TASC, CNR-IOM, Area Science Park, S.S. 14 km 163.5 Basovizza, I-34149 Trieste, Italy\\
$^3$INO-CNR BEC Center and Dipartimento di Fisica, Universit\'a di Trento, I-38123 Povo, Italy}

\begin{abstract}
We demonstrate that intersubband (ISB) polaritons are robust to inhomogeneous effects \jm{originating from the presence of multiple quantum wells (MQWs)}. In a series of samples that exhibit mid-infrared ISB absorption transitions with broadenings varying by a factor of 5 (from 4 meV to 20 meV),  we have observed polariton linewidths always lying in the 4 - 7 meV range only. We have experimentally verified the dominantly inhomogeneous origin of the broadening of the ISB transition, and that the linewidth reduction effect of the polariton modes persists up to room-temperature. This immunity to inhomogeneous broadening is a direct consequence of the coupling of the  large number of ISB oscillators to a single photonic mode. It is a precious tool to gauge \jm{the natural linewidth of the ISB plasmon} , that is otherwise masked \jm{in such MQWs system} , and is also \jm{beneficial} in view of perspective applications such as intersubband polariton lasers.      
\end{abstract}

\maketitle

%

\section{Introduction}
\label{sec:intro}

The mechanisms responsible for broadening of optical transition lines are typically 
classified in two classes: \textit{homogeneous} - related to the dynamics of each 
emitter - and \textit{inhomogeneous} - inherent to the presence of multiple emitters. 
In the textbook example of an atomic gas, the former class contains the radiative 
broadening due to spontaneous emission. The latter class includes Doppler broadening due 
to the wide thermal distribution of velocities, that randomly shifts each emitter resonance \cite{Siegman}. 

A similar physics is found in solid-state systems. For instance, optical transitions in assemblies of quantum dots suffer from a strong inhomogeneous broadening \cite{Gammon}. This comes on top of the radiative 
linewidth and decoherence due to interaction with other degrees of freedom such 
as phonons.
As a first step towards observing the natural linewidth {of quantum dots}, the inhomogeneous broadening due to the slightly different sizes and shapes of the {various dots} can be overcome by restricting to a single object~\cite{footnote}.


%
In the last decades, ISB  transitions in semiconductor quantum wells  have been attracting a growing interest from the fundamental and applied physics community. This is due to their very strong coupling to the electromagnetic 
field, and their easily tunable emission/absorption frequency across a wide spectral range (from terahertz up to mid-infrared frequencies), where few other compact emitters are available. As the frequency of the ISB transition is dependent on the QW thickness \cite{Liu}, and on the doping level \cite{Ando}, slight fabrication inhomogeneities are a major source of broadening of ISB plasmons in MQWs systems. 

\iac{In addition to this, even in the single QW case the physics of ISB transitions is complicated by the homogeneous and inhomogeneous broadening mechanisms that affect the transition linewidth.} On one hand, the homogeneous linewidth is governed by the emission rate of photons and phonons, and \iac{by} the in-plane disorder which opens \iac{additional} decay channels for ISB plasmons. On the other-hand, the non-parabolicity of the electronic bands and the in-plane spatial variations of the QW doping level and thickness are responsible for inhomogeneous broadening  \cite{Liu,Dupont,Campman,Ando,Warburton,khurgin}. \iac{Since the mid '90's this physics has been widely studied for interband excitons in the IR domain}, where several peculiar line-narrowing effects~\cite{whittaker,savona,Ell,Whittaker2}  can be at play to \iac{reduce the transition linewidth} in strong light-matter coupling regimes (see also a review in Ref.~\onlinecite{Liti}). 


\iac{In the ISB context, all these} broadening mechanisms can have dramatic consequences for fundamental studies 
of many body effects in 2D electronic plasmas \cite{Ando}, 
but also for optoelectronic devices. For instance, the threshold current density 
of quantum cascade (QC) lasers is affected by the inhomogeneous broadening 
of the material gain \cite{faist} and can only be partially 
mitigated in the so called broadband QC lasers  \cite{Gmachl,Rosch}. 


\iac{In the last decade, exciting new avenues are being opened by} devices operating in the strong light-matter coupling 
regime, where ISB transitions coupled to a cavity photon mode form new bosonic 
excitations of mixed nature called \textit{ISB polaritons}. 
Such a regime is at the heart of  intriguing proposed applications to low-threshold 
lasers \cite{liberato,colombelli2} and quantum photon sources in the 
far-infrared \cite{ciuti2}. Transferring into the realm of polaritons the design flexibility of ISB transitions - the key ingredient for QC lasers - has been one of the motivations behind the development 
of ISB polaritonics \cite{Dini,Colombelli1}. Still, most of these applications require a narrow polariton
linewidth, which may seem incompatible with the wide absorption linewidth typically observed in devices containing a relatively large number of QWs.

In this {article}, we experimentally show how the strong coupling regime permits to 
largely suppress the inhomogeneous broadening originating from the presence of a large number of QWs. The underlying mechanism is originally discussed in the atomic cavity-quantum electro-dynamics (QED) context \cite{Dalibard,ningyuan}, and it was theoretically translated to the solid-state context in Ref. \onlinecite{houdre}. 
The core idea is that by coupling the \jm {oscillators} to a single cavity mode, one effectively 
singles out a single state out of the inhomogeneous broadened spectral distribution. 
The resulting linewidth reduction turns out to be quantitatively important and 
can \jm {be beneficial} in the development of ISB polaritonics. 

%
%
\begin{figure}
\includegraphics[trim=0.5cm 0cm 0cm 0cm, scale=1.25]{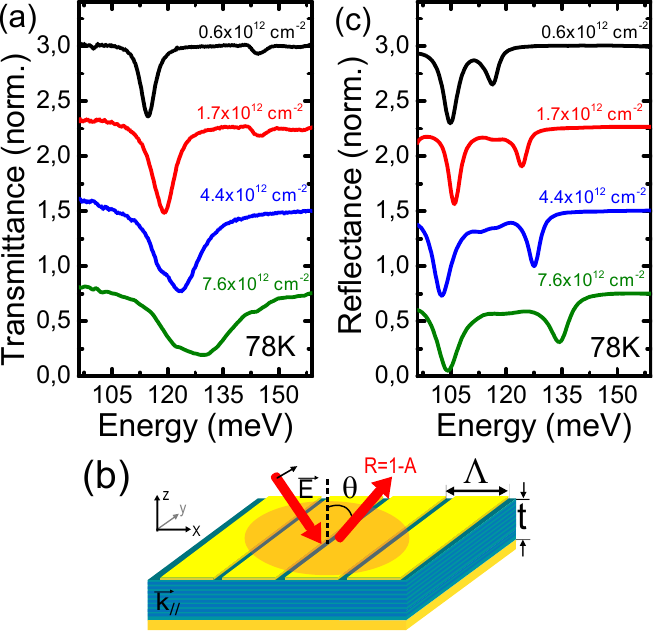}
\caption{(a) Transmission measurements of the 4 studied samples performed at 78 K in a 45$^o$, multipass waveguide configuration. The data are offset for clarity.  (b) Schematic of the polaritonic device and experimental probing conditions. (c) Reflectivity measurements performed on the 4 polaritonic devices at 15$^o$ incidence angle and at 78 K. The reflectivity dips correspond to the lower and upper ISB polaritons. }
\end{figure}
\begin{figure}
\includegraphics[trim=0.75cm 0cm 0cm 0cm, scale=0.31]{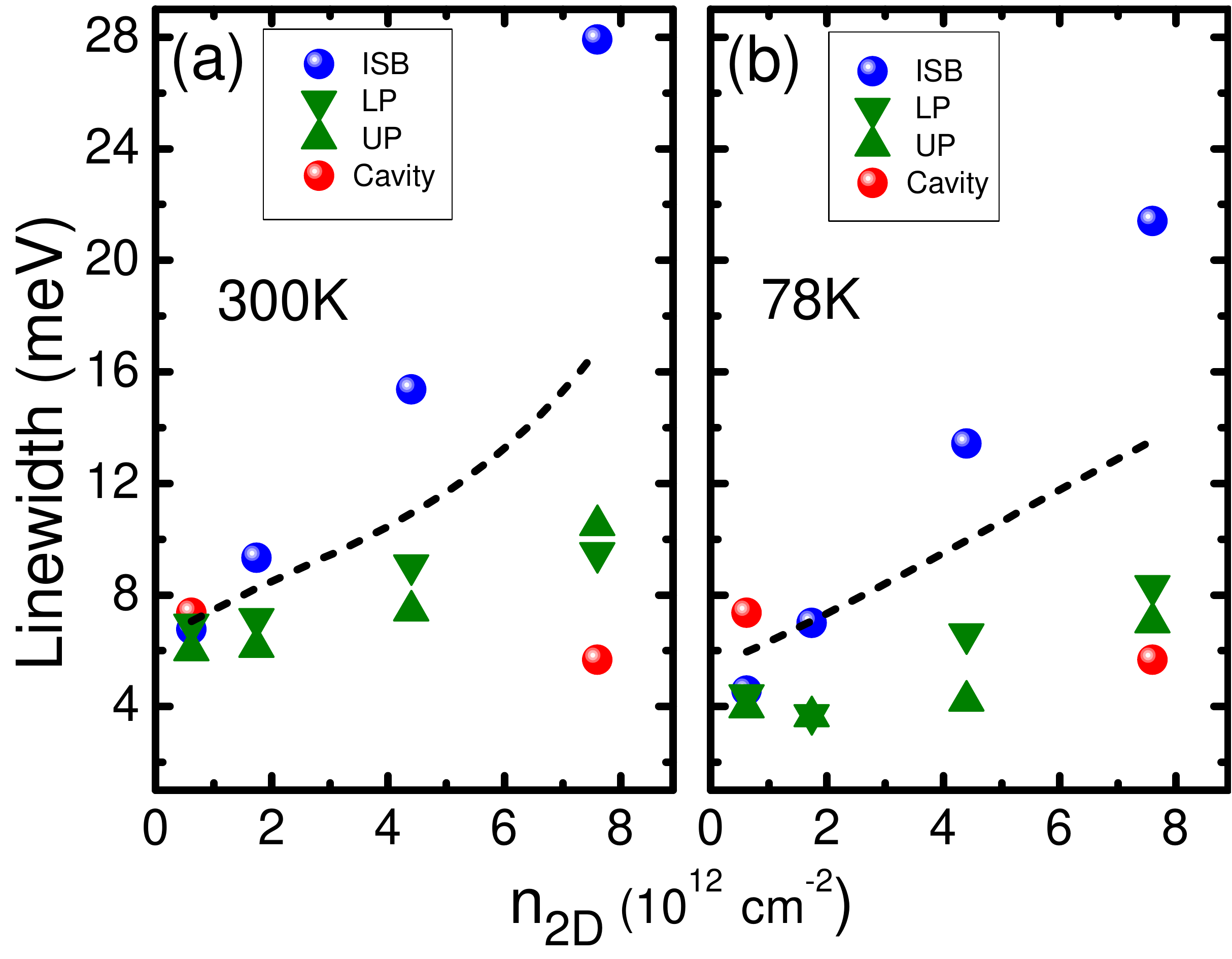}
\caption{Summary of the experimental linewidth of ISB transitions (blue dots), empty grating resonator (red dots), upper polaritons (green up triangles) and lower polaritons (green down triangles). Panel (a) reports the results at 78 K. Panel (b) at 300 K. {The dashed 
lines in the two panels represent the average of cavity and ISB transition linewidths, (${\Gamma}\textsubscript{ISB}+{\Gamma}\textsubscript{cav})/2$ 
that corresponds to the \textit{expected} polariton linewidth at zero detuning, 
i.e. when the Hopfield coefficients are equal to 0.5.}}
\end{figure}

In a series of samples with increasing absorption linewidth, we measure polariton linewidths consistently 
much narrower than the average of cavity mode and ISB plasmon taken independently. 
While hints of this phenomenon have been observed in Ref. \onlinecite{Murphy}, here we provide a full and quantitative characterization of it.  We experimentally demonstrate the dominantly inhomogeneous origin of the ISB plasmon absorption linewidth \jm{in MQWs system}, and we quantitatively explain our observations using a theoretical model inspired from the seminal work Ref. \onlinecite{houdre}.

\section{The experiments}
\label{sec:expts}

The four samples investigated are GaAs/AlGaAs multiple quantum-wells exhibiting 
ISB absorption resonances around 10 \ensuremath{\mu}m wavelength. They have been 
grown by molecular beam epitaxy, and consist in 36-period repetitions of 8.3-nm-thick 
GaAs QWs separated by 20-nm-thick Al$_{0.3}$Ga$_{0.7}$As barriers. The substrate is undoped GaAs, and \ensuremath{\delta}\ensuremath{-}doping 
is introduced in the center of the barriers. The 4 samples differ in the nominal 
modulation doping level \textit{n}\textsubscript{\textit{2D}}: 0.6$\times 10^{12}$ cm\textsuperscript{-2} 
(HM3821), 1.7$\times 10^{12}$ cm\textsuperscript{-2} (HM3820), 4.4$\times 10^{12}$ cm\textsuperscript{-2} 
(HM3872), and 7.6$\times 10^{12}$ cm\textsuperscript{-2} (HM3875).

All the absorption measurements are done using a Fourier transform infrared spectrometer (FTIR) equipped with a Globar thermal source and a deuterated triglycine sulfate (DTGS) detector operating at room temperature. The samples are shaped in multi-pass waveguide configuration and mounted on a continuous flow cryostat having anti-reflection coated Zinc-Selenide (ZnSe) windows implemented on the shroud. Resolution was set at 0.5 meV \iac{(4 cm\textsuperscript{-1})} for all the measurements and the polarization of the incoming radiation is selected with a wire grid holographic Thallium Bromoiodide (KRS-5) in order to get \iac{onto the sample the desired electric-field projection in the direction perpendicular to the growth plane.}

The low temperature (78 K) results are reported in Fig. 1a (results at 300 K are in Supplemental Information~\cite{SM}, Fig. S1). The peak absorption frequency blue-shifts with increasing doping because 
of the depolarization shift\cite{Ando}. Most notably, 
the ISB transitions dramatically broadens with increasing doping.

We have then inserted the four samples in dispersive, grating-based metal-dielectric-metal 
resonators - as sketched in Fig. 1b - following the procedure described in Ref.\onlinecite{manceau}. 
The tight electromagnetic confinement induced by the two metallic surfaces places 
the system in the regime of strong light-matter coupling, and ISB polaritons can 
be probed with surface-reflectivity measurements. \jm{The cryostat is placed within a fixed angle reflection unit (15 degree) inserted within the FTIR and the beam size is selected to cover the entire sample surface (2.5 mm x 2.5 mm). Care is taken to suppress any light that would impinge outside the sample surface.}

Figure~1c reports the reflectivity measurements on all the samples at 78 K and  15\textsuperscript{o} 
incidence. For each sample, we have adjusted the grating period $\Lambda$ (filling factor 
is kept constant at 80\%) in order to obtain the minimum polaritonic splitting 
at \ensuremath{\theta}\ensuremath{\sim}35\textsuperscript{\ensuremath{o}}. 
The measurements reveal the two polaritonic modes as reflectivity dips, whose energy 
distance - the Rabi splitting - increases roughly as the square root of the doping 
(Fig. S2 of Supplemental Information~\cite{SM}), i.e. proportionally to the electronic plasma frequency \label{HHFieldmarkHH367H1262489299}\label{HHFieldmarkHH366H1262489299} \cite{ciuti1,todorov}. 
\begin{figure}
\includegraphics[trim=0.5cm 0cm 0cm 0cm, scale=0.3]{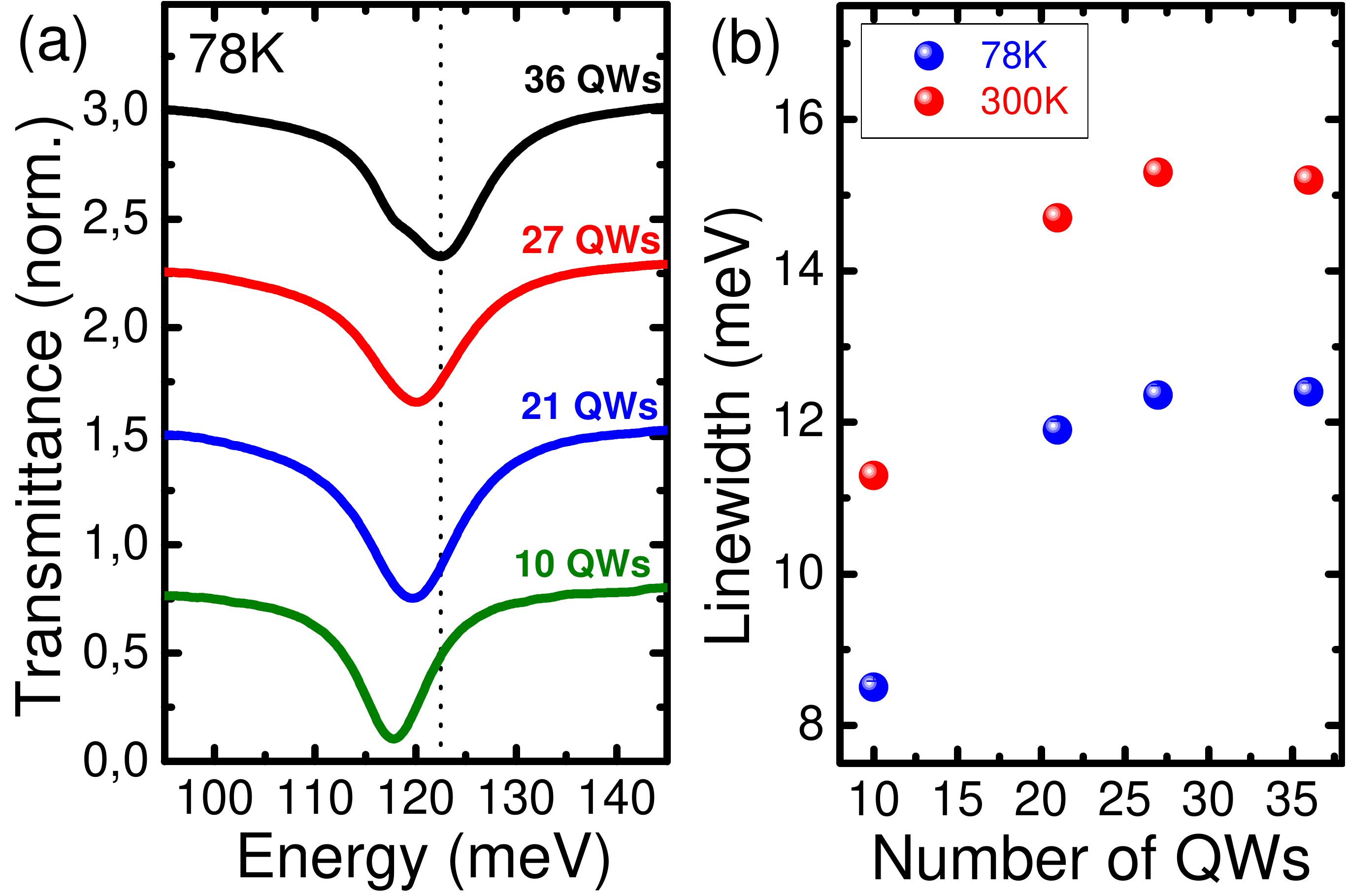}
\caption{(a) Transmission measurements at 78 K of sample HM3872 after removal of 0, 9, 15, and 26 wells respectively. The data are stacked for clarity and the number of remaining QW is reported for each measurement. \jm{The dotted line serves as guide to the eye to mark the central energy of the 36 QWs sample.}(b) Extracted linewidths, using a Voigt fitting procedure, as a function of the number of remaining QWs. }
\end{figure}

The important observation concerns instead the polariton linewidths that appear 
insensitive to the massive broadening of the bare ISB transition with doping. Fig.~2 summarizes the linewidths of ISB absorption transitions (\ensuremath{\Gamma}\textsubscript{ISB}, 
blue dots), upper and lower polaritons (\ensuremath{\Gamma}\textsubscript{UP} and 
\ensuremath{\Gamma}\textsubscript{LP}, green triangles), and bare cavity resonators 
(\ensuremath{\Gamma}\textsubscript{cav}, red dots) at 78 K and 300 K. The dashed 
line represents the average of cavity and ISB transition linewidths, (\ensuremath{\Gamma}\textsubscript{ISB}+\ensuremath{\Gamma}\textsubscript{cav})/2, 
that corresponds to the \textit{expected }polariton linewidth at zero detuning, 
i.e. when the Hopfield coefficients are equal to 0.5 \label{HHFieldmarkHH412H1262489299} \cite{colombelli2}. 
The experimental evidence is instead that \textit{both }\ensuremath{\Gamma}\textsubscript{UP} 
and \ensuremath{\Gamma}\textsubscript{LP  }are much smaller than the expected average 
cavity and ISB transition linewidths. The phenomenon is particularly evident at 
78 K and is not compatible with the expected polariton linewidth being the average of \ensuremath{\Gamma}\textsubscript{ISB} 
and \ensuremath{\Gamma}\textsubscript{cav }\textit{weighted }by the Hopfield coefficients.

{As a final piece of experimental information,}
we demonstrated the dominantly inhomogeneous origin of the bare 
ISB transition broadening observed in Fig.~1a in the absence of the cavity. The 
presence of a low energy shoulder in the absorption spectra of the two highest 
doped samples is already a strong indication. To gain further insight, we measured 
the ISB absorption of three pieces of sample HM3872 (\textit{n}\textsubscript{\textit{2D}}=4.4$\times 10^{12}$ cm\textsuperscript{-2}) 
after removal by sulphuric-acid-based wet chemical etch of 9, 15, and 26 wells 
respectively. The low-temperature (T=78 K) multipass waveguide absorption measurements 
are presented in Fig. 3a: the peak absorption energy red-shifts and the low-energy 
shoulder disappears, thus proving that the 36 QWs composing the sample are not 
all identical. This leads to a linewidth reduction, at both 78 K and room temperature, 
when a reduced number of QWs is probed (Fig. 3b). These measurements prove that 
a large fraction of the ISB linewidth in these samples indeed stems from inhomogeneous 
mechanisms {due to the different parameters of the various wells}.

\section{The theoretical model}
\label{sec:theory}

\iac{Based on this experimental input and taking inspiration from the seminal prediction made in Ref.\onlinecite{houdre}, 
we conjecture that ISB polaritons are not affected by inhomogeneous broadening 
of the MQWs bare ISB transition, but only by the homogeneous one. To substantiate our claim, we generalize the temporal coupled-mode theory of Ref.\onlinecite{manceau} by including the spatial periodicity of the grating-based resonator and the inhomogeneous broadening of the multiple matter oscillators, and we show that this model is able to fully explain in a quantitative way the experimental findings.}

\subsection{Simplified model: planar cavity}

As a first step, we neglect the lateral patterning of the cavity and we build a simplified theory that does not include Bragg scattering processes. This simplification will be useful to fully appreciate the role of the inhomogeneous vs. homogeneous broadenings, before proceeding with the development of a complete theory in the next Subsection.

Under the approximation that the cavity is spatially homogeneous along the $xy$ plane, the in-plane wavevector $\textbf{k}$ is a good quantum number. The system's dynamics can then be written in terms of the following motion equations for the \iac{components at in-plane wavevector  \textbf{k }
of the cavity field and of the ISB oscillator amplitude} in the \textit{j}-th quantum 
well (with $j$=1,...,$N\textsubscript{QW}$) :
\begin{eqnarray}
i\frac{da_{\bm  k}}{dt}&=&\omega^{cav}_{\bm k} a_{\bm k}-i\frac{(\gamma_{rad}+\gamma_{nr})}{2} a_{\bm k}+ \nonumber \\ &+&\Omega\sum_j b_{j,{\bm k}}+E_{inc}(t) \label{eq:a} \\
i\frac{db_{j,{\bm k}}}{dt}&=&\omega^{ISB}_{j} b_{j,{\bm k}}+\Omega a_{\bm k}-i\frac{\gamma_{hom}}{2} b_{j,{\bm k}}. \label{eq:b}
\end{eqnarray}
%
Here,  $\omega^{cav}_{k}$  denotes the cavity mode dispersion, 
and the frequencies $\omega^{ISB}_{j}$ of the ISB transition 
in each well are assumed to be  independent of the in-plane wavevector \textbf{k} and distributed 
around their central frequency $\omega$\textsubscript{ISB} according to a Gaussian 
distribution of standard deviation $\sigma$\textsubscript{inh}.  \iac{Furthermore,} $\Omega$ is the 
Rabi frequency of each ISB plasmon coupling to the cavity mode. $E_{inc}$ 
is the incident field. $\gamma_{rad}$ and $\gamma_{nr}$ are the radiative 
and non-radiative linewidths of the cavity mode, and $\gamma_{hom}$ is the 
homogeneous linewidth of the ISB plasmon, {resulting from all non-radiative decay and decoherence mechanisms due, e.g., to electron scattering on interface roughness~\cite{Liu,Dupont,Campman,Ando,Warburton,khurgin}.} The reflected field results then 
from the interference of the directly reflected incident field and the cavity emission \cite{ciuti2},  
{
\begin{equation}
E_{refl}=E_{inc} - i \gamma_{rad} a_{\bm k}.
\label{eq:refl}
\end{equation}
Reflection spectra are then straightforwardly obtained by inverting the linear set of equations describing the steady-state 
of the motion equations (\ref{eq:a})-(\ref{eq:b}) and inserting the result into (\ref{eq:refl}). An intuitive physical understanding of this model can be summarized as follows along the lines of Ref.\onlinecite{houdre}. 
 
In the absence of inhomogeneous broadening, all wells are identical $\omega^{ISB}_{j}= \omega^{ISB}$, so
the light-matter coupling singles out the fully symmetric combination of 
ISB plasmons \textit {b}$_{B,\bm{k}}$=  $\Sigma_{j}$\textit {b}$_{j,\bm{k}}/\sqrt{N_{QW}}$. This single \textit{bright} combination couples to the cavity mode with a collective Rabi frequency $\Omega_{R}=\sqrt{N_{QW}}\Omega$. Coherent mixing 
of the bright ISB and the cavity modes} gives rise to the ISB polaritons, whose 
linewidth results from a weighted average of the cavity and ISB homogeneous linewidths. 
All other combinations of ISB's remain at $\omega$\textsuperscript{ISB  }but are 
dark and therefore do not appear in the optical spectra. 

The situation is more interesting in the presence of some inhomogeneous broadening. 
{As long as its standard deviation $\sigma$\textsubscript{inh} does not exceed the 
collective $\Omega$\textsubscript{R}, the inhomogeneous broadening of the ISB plasmons is only responsible for a corresponding spectral broadening of the the dark states and a weak mixing of them with the bright states. As a consequence, dark states transform into a wide band of weakly optically active states located in between the polaritons, which however remain spectrally well separated and almost unaffected by the inhomogeneous broadening.} 
Since large $\Omega$\textsubscript{R }values are a peculiar character of ISB polariton 
systems, this mechanism explains why {ISB polaritons} are extremely robust against inhomogeneous 
broadening. 

The {behaviour} suddenly changes when the  \iac{inhomogeneous broadening $\sigma$\textsubscript{inh} becomes comparable to} the collective $\Omega$\textsubscript{R}: in this case, \iac{the energy range over which dark and bright states are mixed by the inhomogeneous broadening} reaches the spectral position of polaritons. Seen in the polariton basis, the magnitude of the mixing terms can cross the polariton gap and \iac{effectively} contaminate the polaritons with the dark states. As a result, the two polaritons lose their character of spectrally isolated states and their 
linewidth suddenly increases washing out the corresponding spectral features.

\subsection{Complete theory including Bragg processes}

While the simplified is able to account for the main features of the inhomogeneous broadening, it does not include the \iac{spatial periodicity of period $a$ of} the top mirror and the subsequent Bragg scattering processes that are responsible for the folding of the photonic and polaritonic bands and the consequent Bragg gaps that open between them.

In this Subsection we develop a complete theory that is able to quantitatively reproduce the experimental results. \iac{Restricting for simplicity our attention to the $k_y=0$ line that is addressed in the experiments, the photon and ISB bands can be labeled by the $k_x$ component along the spatial periodicity, denoted for brevity $k$ and belonging to the first Brillouin zone $k \in [-k_{Br}/2,k_{Br}/2]$ with $k_{Br}=2\pi/a$, and by the band index $n=1,2,3,\ldots$.}

The equation of motion for the amplitudes in the photonic and ISB modes read:
\begin{eqnarray}
	i \frac{da_{n,{ k}}}{dt} &=& \omega^{cav}_{n,{ k}}  a_{n,{ k}} +\Omega \sum_j b_{j,n,{ k}}  + \Omega_{Br} [a_{n+1,{ k}} + a_{n-1,{ k}}]  \nonumber \\ &-&i\frac{\gamma_{nr}}{2} a_{n,{ k}}-i\frac{\gamma_{rad}}{2} \sum_{n'} \sqrt{\eta_{n,{ k}}\eta_{n',{ k}}} a_{n',{ k}} + \nonumber \\
	&+&\sqrt{\eta_{n,{ k}}}\,E_{inc}(t) \label{eq:a_br} \\
        i \frac{db_{j,n,{ k}}}{dt}&= & \omega^{ISB}_j  b_{j,n,{ k}} + \Omega a_{n,{ k}} - i \frac{\gamma_{hom}}{2} b_{j,n,{ k}} \label{eq:b_br} 
 \end{eqnarray}
The momentum-independence of the ISB transitions for wavevectors much smaller than the Fermi momentum of the electron gas reflects in the independence of $\omega^{ISB}_j$ from the in-plane wavevector ${ k}$ and the band index $n$. The inhomogeneous broadening of the ISB of the different wells (labeled by $j$) has the same shape as discussed in the previous section for the simplified model. 

The periodic patterning of the mirror is described by letting the frequencies $\omega^{cav}_{n,{ k}}$ of the photonic bands to follow the folded dispersion
\begin{eqnarray}
\omega^{cav}_{n,{ k}}  &=& \frac{c}{n_0} \left[k +(n-1) \frac{k_{Br}}{2}\right] \;\;\; \textrm{for}\;\;n\;\; \textrm{odd} \\
\omega^{cav}_{n,{ k}}  &=& \frac{c}{n_0} \left[n\frac{k_{Br}}{2}-k\right] \;\;\; \textrm{for}\;\; n\;\; \textrm{even}
\end{eqnarray}
for $k>0$, and a symmetric one for $k<0$. 
The coefficient $n_0$ is the effective refractive index of the cavity, determined by a combination of the bulk material index and the penetration into the mirrors. The amplitude of the Bragg processes opening gaps at the center and at the edges of the Brillouin zone is quantified by $\Omega_{Br}$. 

The $\eta_{n,{ k}}$ coefficients account for the different radiative coupling of each band to the external radiative \iac{modes and are responsible for the peculiar radiative couplings~\cite{fano} that appear in the equations of motion (\ref{eq:a_br}) for the field amplitudes}. In our numerics, we have considered the simplest case where this coupling coefficient vanishes for the lowest \iac{$n=1$} band (which falls out of the light cone), is full and equal to 1 for the $n=2,3$ bands, and then drops to a small value of $0.1$ for the higher bands (to model the reduced large-angle scattering amplitude by the mirror patterning). While the details of the high-$n$ behavior are actually not important to match the experimental results, the light-cone condition on the lowest \iac{$n=1$} band and, even more, the simultaneous and comparable radiative coupling of the $n=2,3$ bands are essential to reproduce the \iac{experimentally observed dependence of the visibility of the different photonic bands on the band index $n$ and on the wavevector $k$.}

Finally, the reflection spectra can be obtained by numerically inverting the linear set of equations describing the steady-state 
of the motion equations (\ref{eq:a_br})-(\ref{eq:b_br}) and inserting the result into the generalized reflection amplitude,
\begin{equation}
E_{refl}=E_{inc} - i \gamma_{rad} \sum_n \sqrt{\eta_{n,{ k}}}\,a_{n,{\bm k}}.
\label{eq:refl_br}
\end{equation}
\iac{This approach straightforwardly lead to reflection spectra such as the ones shown in Fig.4(b,c).}

\subsection{Comparison with experiments}
\begin{figure}
\includegraphics[trim=5.5cm 5cm 0cm 3cm, scale=0.65]{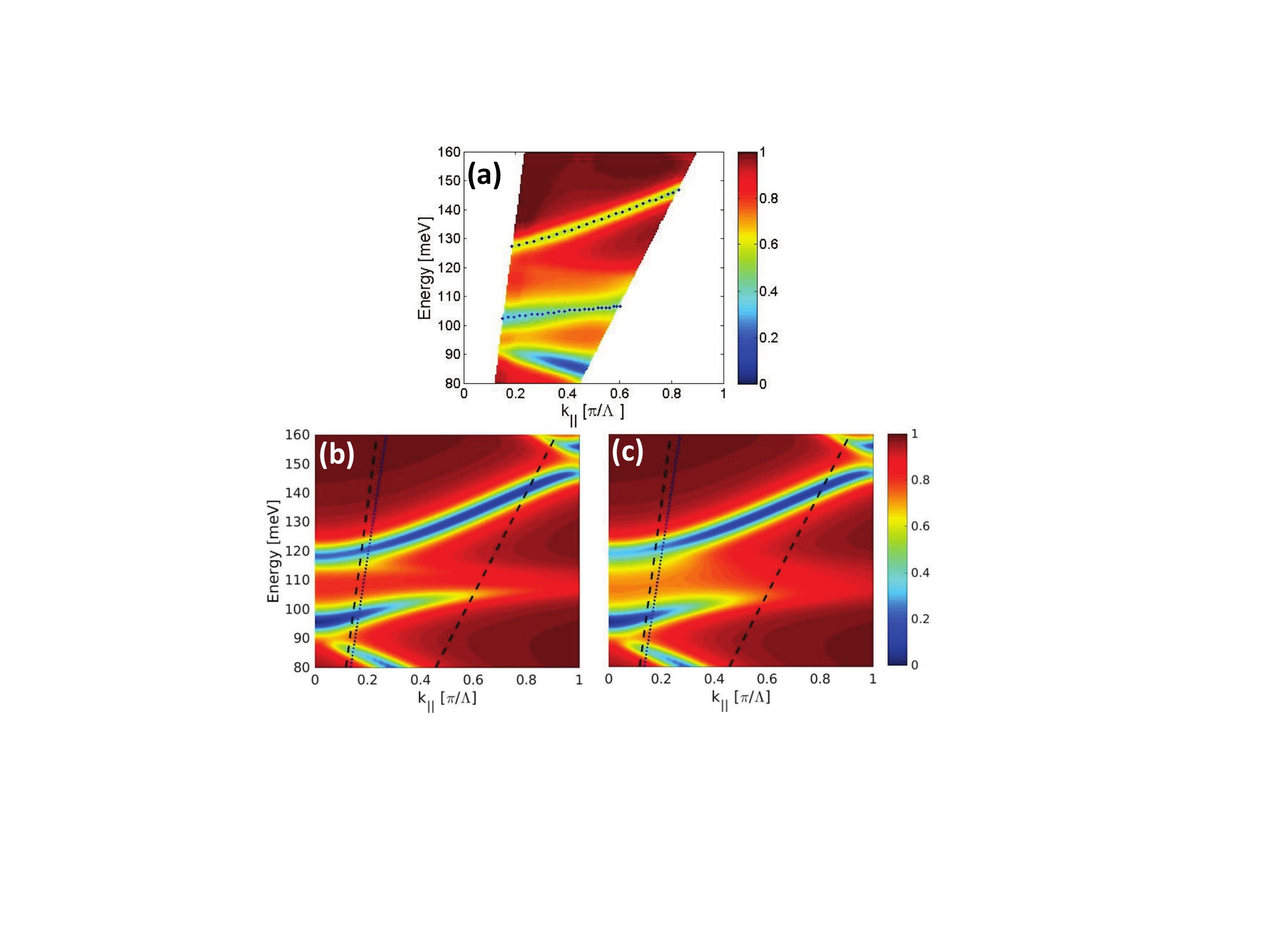}
\caption{(a) Experimental band-diagram of the polariton sample HM3872. The measurements are performed at 300 K. The dispersion is obtained by measuring the zero-order reflectivity of the sample between $\theta=13^o$ and $\theta=61^o$. The reflectivity minima of the polaritonic branches are marked with blue dots.
(b,c) Theoretically calculated reflectivity as a function of the in-plane wavevector and the frequency for no inhomogeneous broadening w$_{inh}$=0 (b) and for the experimentally realistic value $w_{inh}$=12 meV (c). All other parameters are chosen to match the experiment and are given in the text and in the SM. The black dashed lines indicate the edges of the experimentally accessible region. The dotted line indicates the $\theta=15^o$ line at which the spectra in Fig.5 are taken. }
\end{figure}

{We are now going to show how this theoretical model} provides a quantitative 
explanation of the experimental observations. Figure 4(a) reports the experimental 
polaritonic band-structure of sample HM3872 (\textit{n}\textsubscript{\textit{2D}}=4.4$\times 10^{12}$ cm\textsuperscript{\textit{-2}}) 
experimentally acquired at room-temperature. It is obtained by measuring the zero-order 
reflectivity of the sample between \ensuremath{\theta}\ensuremath{=}13\textsuperscript{\ensuremath{o}} 
and \ensuremath{\theta}=61\textsuperscript{o}, and then applying the transformation 
$k=\frac{\omega}{c}\sin\theta$ to obtain the in-plane wave-vector $k_{\parallel}$. \jm{For this measurement, the sample is placed in a motorized angular-reflection unit \iac{located} within the chamber of the FTIR. The entire surface of the sample is probed and care is taken to suppress any light impinging outside the sample surface, especially at high angles where the beam spread with a cos($\theta$) dependency.}
  
The corresponding numerically calculated polaritonic dispersions are shown in Figs.~4(b,c), respectively in the presence and in the absence of an inhomogeneous broadening 
of magnitude $\sigma$\textsubscript{inh}=5.1 meV (that is FWHM $w_{inh}=12$~ 
meV) comparable to the experimental one around the central frequency $\omega^{ISB}=110$~meV. The homogeneous broadening is taken as $\gamma_{hom}$= 6 meV. {The cavity radiative and non-radiative linewidths are $\gamma_{rad}$=2.5 
meV and $\gamma_{nr}$=3.5 meV, respectively, while the effective refractive index is $n_0=3.1$ and the Bragg coupling $\Omega_{Br}=5$~meV. The collective Rabi frequency is taken as $\Omega_R=11\,\textrm{meV}$.}

{The quantitative agreement between the theoretical model and the experimental data is very good: the two strongest lines corresponding 
to the upper and lower polaritons are almost unaffected by the inhomogeneous broadening 
and show linewidths of approximately $\gamma_{LP,UP }$=4 meV each. This 
correctly reproduces the experimental observation of polaritonic linewidths that 
are much smaller than the absorption linewidth of bare ISB transition shown in Fig.1(a). 

The faint features visible in Fig.4(b) within the polariton gap between 105 and 120 
meV are due to the folding of the polaritons by the Bragg periodicity. As it is shown in Fig.4(c), their structure is washed out by the inhomogeneous broadening and \iac{they} transform into an almost structureless band stemming from the weak mixing of the dark states with the bright ones. A weak trace of these features is anyway still visible in the experimental band-diagram of Fig.4(a) and in the reflection spectra of Fig.1(c). }
\begin{figure}
\includegraphics[trim=0cm 0cm 0cm 1cm, scale=0.3]{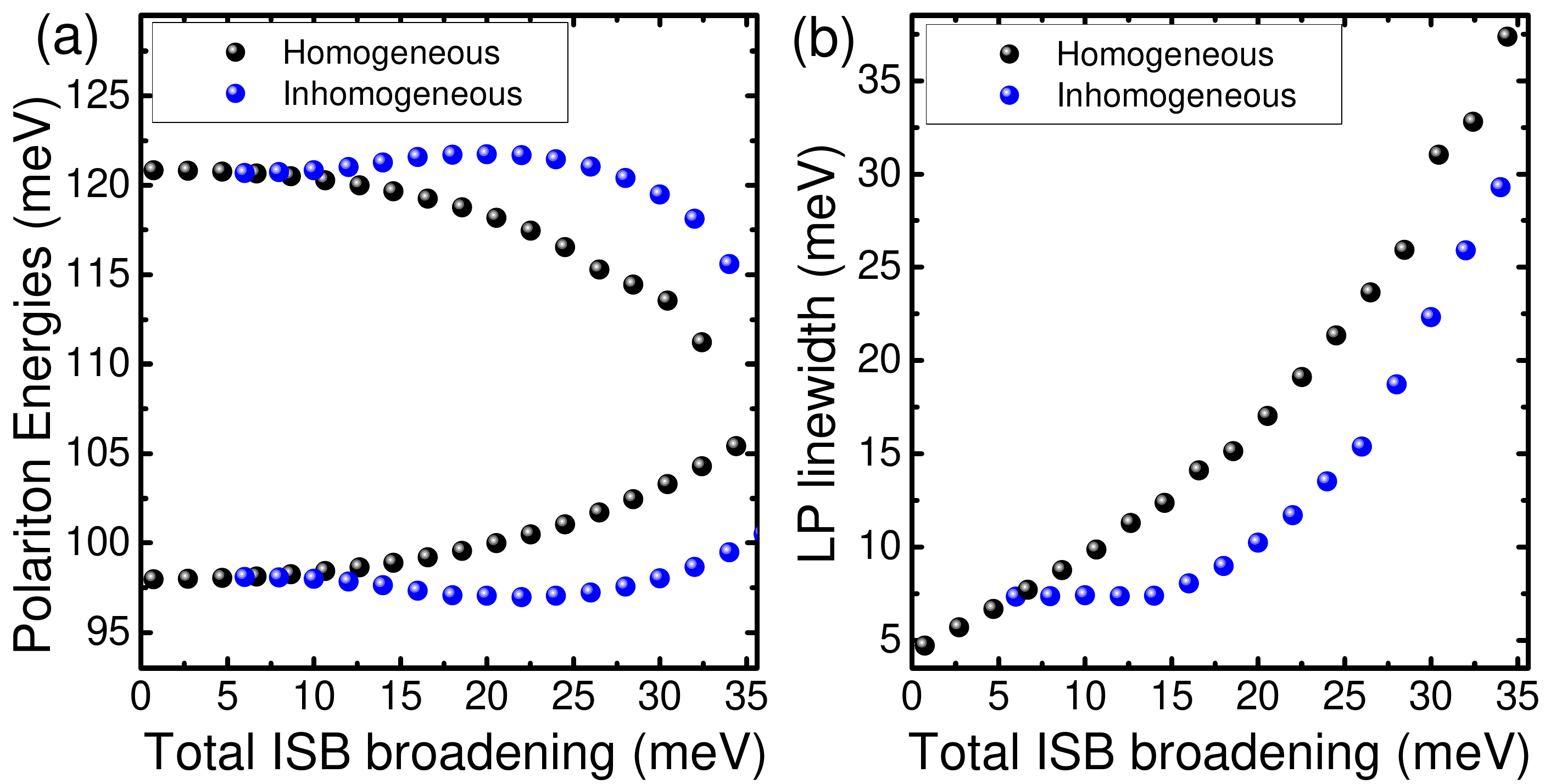}
\caption{Polaritons peaks positions and lower polariton linewidth as extracted from theoretically calculated reflectivity spectra at 15$^o$ for growing values of the total ISB broadening. Black {dots} correspond to the homogeneous case, while blue dots to the inhomogeneous one. \iac{In this latter case,} the homogeneous broadening is kept constant at an experimental value of $\gamma_{hom}$= 6 meV).
In (a), the polariton peaks positions are reported. A rapid reduction of the splitting is observed in the homogeneous case, while it is absent in the inhomogeneous one.
In (b), a similar effect is observed in the linewidth of the lower polariton state.  }
\end{figure}

To corroborate the findings, and to better illustrate the consequences of this 
phenomenon, the different impact of inhomogeneous or homogeneous broadenings is theoretically illustrated 
in Fig.5. 
The two (a,b) panels respectively show the polariton peak positions and the linewidth of the LP 
mode, extracted from the reflection spectra calculated at a given incidence 
angle of 15\textsuperscript{o}, as a function of the total ISB linewidth $\gamma_{tot}$.
In the homogeneous case, $\gamma_{tot} = \gamma_{hom}$.
In the inhomogeneous case,  
{ $\gamma_{tot} = \gamma_{hom}+\sigma_{inh}$,} and
the homogeneous contribution is kept constant at a value  $\gamma_{hom}$=  6 meV.

Fig.~5(a) shows that the Rabi splitting immediately starts to quench with increasing homogenous broadening, 
while it is much more stable against an equivalent increase of inhomogeneous one. 
For instance, a purely homogeneous ISB linewdith of 22 meV significantly reduces the Rabi splitting to $2\Omega_{Rabi}=17$ meV, at the
onset of weak coupling.
On the other hand, {in the inhomogeneous case the Rabi splitting remains stable around a value $2\Omega_{Rabi}\approx 25$ meV well within the strong coupling regime and is even slightly reinforced by the spectral broadening as compared to the purely homogeneous case.}

Fig.~5(b) highlights the different effect of homogeneous and inhomogeneous broadening on the polariton linewdith.
In the homogeneous case, the LP FWHM increases approximately linearly with increasing ISB broadening, as expected.
In the inhomogeneous case, instead, the LP FWHM is essentially unaffected until $\gamma_{tot}$ is of the order of the 
initial vacuum field Rabi splitting, $\approx$20 meV. At that point it starts increasing, but --importantly-- it is always smaller than in the homogeneous case.

These data highlight the qualitative difference between the two cases: the homogeneous broadening  is responsible for a rapid reduction of the Rabi splitting with simultaneous 
broadening of the polaritonic states, which gradually morph into a single bare cavity photon line as typical of weak coupling. On the other hand, the inhomogeneous broadening does not initially affect the Rabi splitting nor the polariton linewidths, which remain approximately 
constant. Only when the inhomogeneously broadened ISB transition exceeds the polariton splitting, the  polaritonic states suddenly collapse into a single broad photon line.

\section{Conclusions}
\label{sec:conclu}

In conclusion, we have experimentally demonstrated how the strong light-matter 
coupling regime permits to strongly reduce the inhomogeneous broadening of ISB transitions in a multiple semiconductor QW system, at least its component stemming from the presence of a large number of slightly different QWs.
{The mechanism underlying the observed line narrowing effect is} a direct consequence of (i) the coupling  between a large number of oscillators and a single photonic mode and (ii) the elevated coupling constants typical of ISB polariton systems.  

In addition to offering a  clear illustration of a general physical mechanism, the large line narrowing achievable makes this result an important  step in the direction of extending solid-state quantum optical experiments to intermediate  wavelengths between the visible/IR range of interband devices and the microwaves  of circuit-QED devices. \rc{Furthermore, the elucidated mechanism could be useful to disentangle the homogeneous and inhomogeneous contributions of in-plane disorder to the ISB transition linewidth in single quantum wells.}
\\ \indent

We thank F. Julien, C. Ciuti, G. C. La Rocca, S. De Liberato for useful discussions. This work 
was partly supported by the French RENATECH network. We acknowledge financial support 
from European Union FET-Open grant MIR-BOSE 737017.

\end{document}